\documentclass[prb,twocolumn,amsmath,amssymb,floats,floatfix,prl,aps,superscriptaddress,showkeys]{revtex4}

\usepackage{graphicx}% Include figure files
\usepackage{epsfig}
\usepackage{dcolumn}% Align table columns on decimal point
\usepackage{bm}% bold math
\usepackage{times}

\begin{document}

\title{\bf Rabi oscillations at the exceptional point in anti-parity-time symmetric diffusive systems}
\author{Gabriel Gonz\'alez\thanks{C\'atedra CONACYT, Universidad Aut\'onoma de San Luis Potos\'i, San Luis Potos\'i, 78000 MEXICO.
E-mail:~{\tt gabriel.gonzalez@uaslp.mx}}}
\date{\today}

\begin{abstract}
The motivation for this theoretical paper comes from recent experiments of a heat transfer system of two thermally coupled rings rotating in opposite directions with equal angular velocities that present anti-parity-time (APT) symmetry. The theoretical model predicted a rest-to-motion temperature distribution phase transition during the symmetry breaking for a particular rotation speed. In this work we show that the system exhibits a parity-time ($\mathcal{PT}$) phase transition at the exceptional point in which eigenvalues and eigenvectors of the corresponding non-Hermitian Hamiltonian coalesce. We analytically solve the heat diffusive system at the exceptional point and show that one can pass through the phase transition that separates the unbroken and broken phases by changing the radii of the rings. In the case of unbroken $\mathcal{PT}$ symmetry the temperature profiles exhibit damped Rabi oscillations at the exceptional point. Our results unveils the behavior of the system at the exceptional point in heat diffusive systems.
\end{abstract}
\maketitle

A closed or conservative system evolves according to a Hermitian Hamiltonian in contrast with open or non conservative systems which are described by non-Hermitian Hamiltonians. There are a special class of non-Hermitian systems in which the energy exchange between the system and the environment is balanced. The entire balanced system exhibits a symmetry called $\mathcal{PT}$ symmetry where the symbol $\mathcal{P}$ stands for parity and interchanges the gain and loss components of the total system and $\mathcal{T}$ represents the operation of time reversal and has the effect of turning a system with loss into a system with gain and viceversa.\cite{bender, bender1, bender2}\\ 
Non-Hermitian $\mathcal{PT}$ symmetric systems can exhibit a rich and unexpected behavior and have broad applications in classical and quantum physics.\cite{bender3,bender4,bender5,bender6} $\mathcal{PT}$ symmetric systems have been intensively studied in optics in which many intriguing phenomena haven been experimentally confirmed and has led to the development of new ways of controlling light propagation.\cite{guo, ruter, chang, wimmer}\\
Recently, anti-$\mathcal{PT}$ (APT) symmetric systems have attracted a lot of attention because they exhibit noteworthy effects different from the $\mathcal{PT}$ counterpart. An APT symmetric Hamiltonian can be defined in terms of a $\mathcal{PT}$ symmetric Hamiltonian by $H^{(APT)}=\pm iH^{(PT)}$, but physically it is really difficult to implement it in the laboratory since it requires the coupling between the two subsystems to be a purely imaginary value, in contrast with the $\mathcal{PT}$ systems which requires a real coupling. Anti-PT symmetry has been demonstrated by using dissipatively coupled atomic beams,\cite{peng} cold atoms,\cite{jiang} electrical circuits,\cite{choi} and optical devices.\cite{zhang,lic,zhao} These breakthroughs have initiated the
field of exploring unique APT effects. More recently, Li {\it et al} reported the experimental realization of an APT symmetric diffusive system in Ref.\cite{X}. The system investigated in Ref.\cite{X} is depicted in Fig. 1 and consists of two identical solid rings with inner and outer radius given by $R$ and $R+\delta R$, respectively. The thickness is $b$. The upper ring is rotating with angular velocity $\omega_1$, while the lower ring is rotating with angular velocity $\omega_2=-\omega_1$. There is an interface of thickness $d$ and thermal conductivity $k_i$ between the two rings. The temperature distribution along the inner edges of the upper and lower rings is given by the following diffusion coupled partial differential equations
\begin{align}
\frac{\partial T_1}{\partial t}=D\frac{\partial^2 T_1}{\partial x^2}-v\frac{\partial T_1}{\partial x}+h_c(T_2-T_1) \nonumber \\
\frac{\partial T_2}{\partial t}=D\frac{\partial^2 T_2}{\partial x^2}+v\frac{\partial T_2}{\partial x}+h_c(T_1-T_2)
\label{eq03}
\end{align}
\begin{figure}[t]
  \begin{center}
    \begin{tabular}{cc}
      \resizebox{40mm}{!}{\includegraphics[trim=0 -3cm 0 0]{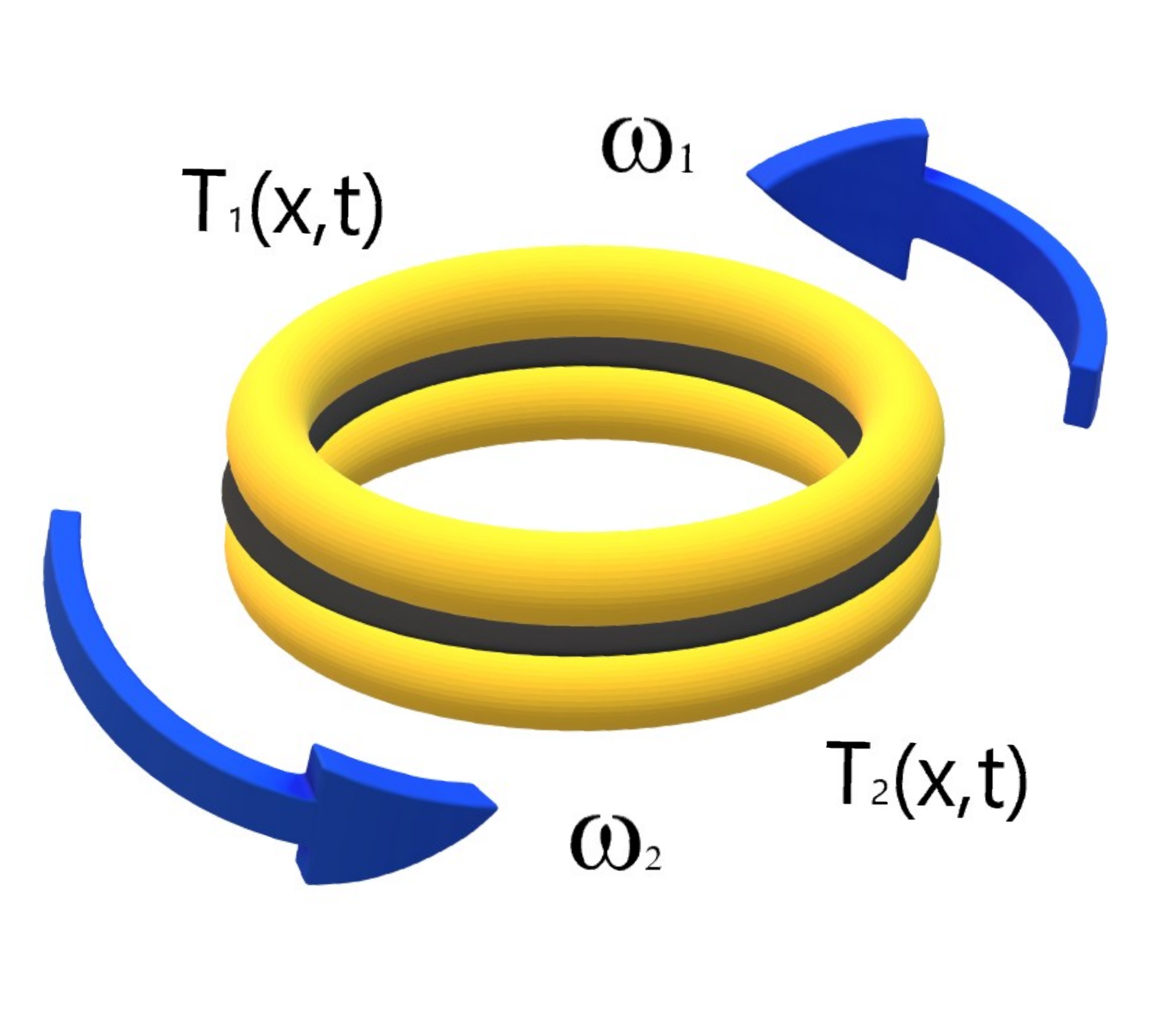}} &
      \resizebox{40mm}{!}{\includegraphics{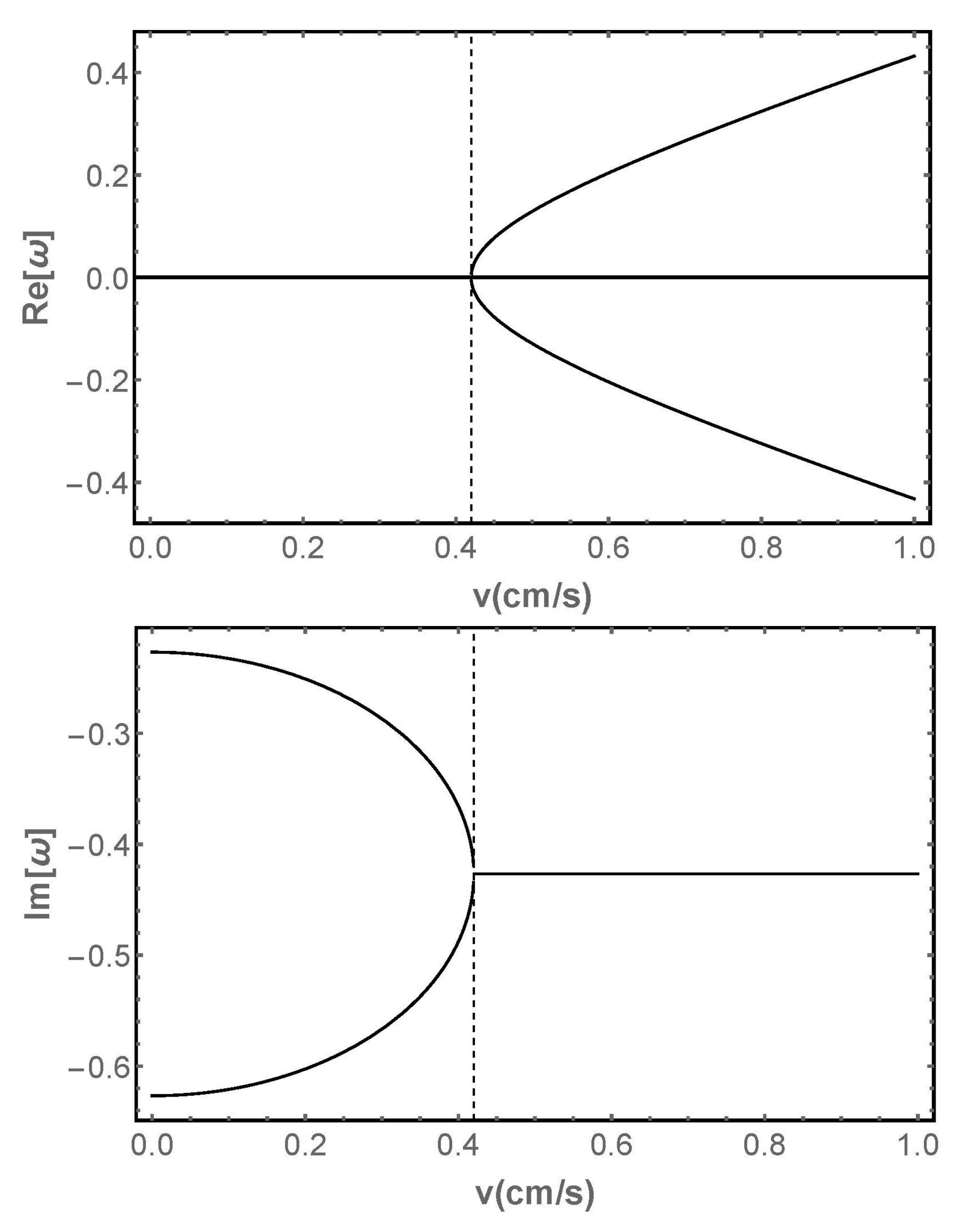}} \\
    \multicolumn{1}{c}{\mbox{\bf (a)}} &
\multicolumn{1}{c}{\mbox{\bf (b)}} \\
    \end{tabular}
    \caption{The figure (a) shows two identical rotating rings with equal but opposite angular velocities joined together by a stationary intermediate layer and (b) the imaginary and real parts of the eigenfrequencies as a function of the tangential velocity where the dotted line represents the exceptional point $v_{EP}=h_c/\kappa$. }
\label{BPBT}
  \end{center}
\label{fig1}
\end{figure}
where $x$ is the coordinate along each edge, $D=k/\rho c$ is the diffusivity, $v$ is the tangential velocity in the inner edge of the rings,$h_c=h/\rho cb$ is the rate of heat exchange coupling, $\rho$ is the density, $c$ is the heat capacity and $h=k_i/d$ is a coefficient that represent the heat exchange between the two rings. Using plane wave solutions, i.e. $T_i=A_ie^{i(\kappa x-\omega t)}$, the system given in Eq. (\ref{eq03}) can be cast into an APT symmetric Hamiltonian given by 
\begin{equation}
H^{(APT)}=
\begin{pmatrix}
-i(\kappa^2D+h_c)+\kappa v & ih_c \\
ih_c & -i(\kappa^2D+h_c)-\kappa v
\end{pmatrix}
\label{eq03a}
\end{equation}
where $\kappa$ is the wave number and $\omega$ are the eigenvalues of the APT Hamiltonian which are given by
\begin{equation}
\label{eq03b}
\omega_{\pm}=-i\left[(\kappa^2D+h_c)\pm\sqrt{h_c^2-\kappa^2v^2}\right].
\end{equation}
The exceptional point where the two eigenvectors coalesce is when $v_{EP}^2=h_c^2/\kappa^2$, i.e. $\omega_+=\omega_-$. The sudden collapse of the eigenvectors and eigenvalues at the exceptional point leads to an abrupt reduction in dimensionality.
Many of the interesting properties of non Hermitian systems are found at or close to the exceptional point which have led to many novel and exotic phenomena. Exceptional points are currently the subject of many interesting and counter-intuitive phenomena associated with them such as topological mode switching,\cite{topo1,topo2} reflection and transmission,\cite{refle1,refle2,refle3} instrinsic single-mode lasing\cite{intri1,intri2} and coherent perfect absorption.\cite{coher} \\
In this work we study the APT symmetric diffusive system given by Eq. (\ref{eq03}) when $v=v_{EP}$ and show that the system behaves as a pair of coupled linear oscillators one with gain and the other one with loss. The noteworthy feature of the exceptional point $v_{EP}$ is that it exhibits damped Rabi oscillations in the unbroken $\mathcal{PT}$ phase transition that depends on the radii of the rotating rings. We obtain the analytical temperature distribution of each ring at the exceptional point and obtain the conditions that have to be fulfill in order for the system to be in equilibrium. We start our investigation by making the following variables change in Eq. (\ref{eq03}): $\tau=h_ct$, $z=\sqrt{h_c(\lambda-1)/D}x$ where $\lambda>1$ is an auxiliary constant to be determined and $\Delta T=T_i-T_0$ where $T_0$ is a reference temperature. Rewriting Eq. (\ref{eq03}) in terms of the new variables we have
\begin{widetext}
\begin{align}
h_c\frac{\partial \Delta T_1}{\partial \tau}=h_c(\lambda-1)\frac{\partial^2 \Delta T_1}{\partial z^2}-v_{EP}\sqrt{\frac{h_c(\lambda-1)}{D}}\frac{\partial \Delta T_1}{\partial z}+h_c(\Delta T_2-\Delta T_1) \nonumber \\
h_c\frac{\partial \Delta T_2}{\partial \tau}=h_c(\lambda-1)\frac{\partial^2 \Delta T_2}{\partial x^2}+v_{EP}\sqrt{\frac{h_c(\lambda-1)}{D}}\frac{\partial \Delta T_2}{\partial z}+h_c(\Delta T_1-\Delta T_2)
\label{eq04}
\end{align}
\end{widetext}
Looking for solutions of the form $\Delta T_i=e^{-\lambda \tau}f_i(z)$ in Eq. (\ref{eq04}) we end up with the following system of coupled ordinary differential equations
\begin{align}
\frac{d^2f_1}{dz^2}-\frac{v_{EP}}{\sqrt{Dh_c(\lambda-1)}}\frac{df_1}{dz}+f_1+\frac{1}{\lambda-1}f_2=0 \nonumber \\
\frac{d^2f_2}{dz^2}+\frac{v_{EP}}{\sqrt{Dh_c(\lambda-1)}}\frac{df_2}{dz}+f_2+\frac{1}{\lambda-1}f_1=0.
\label{eq05}
\end{align}
Inspection of Eqs.(\ref{eq05}) reveals that they are invariant under combined parity, i.e. $f_1\leftrightarrow f_2$, and time reversal $t\rightarrow -t$ transformation. To solve the system of equations analytically we first differentiate one of the equations and then use the other equation to eliminate $f_2$ in order to get the following fourth order differential equation 
\begin{equation}
\bigg(\frac{d^4}{dz^4}+\left(2-\frac{\epsilon v^2_{EP}}{Dh_c}\right)\frac{d^2}{dz^2}+(1-\epsilon^2)\bigg)f_1(z)=0.
\label{eq06}
\end{equation}
where $\epsilon=1/(\lambda-1)$. 
By assuming a solution of the form $f_1(z)\propto\cosh(\chi z)$ for Eq. (\ref{eq06}) we get the following condition over $\chi$:
\begin{equation}
\chi^4+(2-a^2)\chi^2+(1-\epsilon^2)=0,
\label{eq08}
\end{equation}
where $a^2=\epsilon v_{EP}^2/Dh_c$.
The solution of Eq. (\ref{eq08}) is given by
\begin{equation}
\chi^2=\frac{1}{2}\left(a^2-2\pm\sqrt{a^4-4a^2+4\epsilon^2}\right).
\label{eq09}
\end{equation}
In order to have an oscillatory behavior we must demand that $\chi^2<0$, which implies that
\begin{itemize}
 \item[(i)] $a^4-4a^2+4\epsilon^2 > 0$ \\
 \item[(ii)] $a^2-2+\sqrt{a^4-4a^2+4\epsilon^2} < 0$.
\end{itemize}
Condition (ii) gives $\epsilon<1$ and condition (i) gives 
\begin{equation}
a<a_{crit}=2\left(1-\sqrt{1-\epsilon^2}\right).
\label{eq10}
\end{equation}
If $\epsilon<1$ and $a<a_{crit}$ we get the following oscillatory solution for $f_1$
\begin{equation}
f_1(z)=A_1\cos\left(\chi_1 z\right)+B_1\cos\left(\chi_2 z\right),
\label{eq10a}
\end{equation}
where $\chi_{1,2}=\sqrt{|\chi_{\pm}^2|}$ and $A_{1}$ and $B_1$ are constants to be determined. In order to obtain the value of $\epsilon$ we must consider the periodicty of $f_1(z)$, i.e. $f(0)=f(2\pi R\sqrt{h_c/D\epsilon})$, which gives us the following conditions 
\begin{equation}
\chi_{1,2}R\sqrt{\frac{h_c}{D\epsilon}}=n, 
\label{eq10b}
\end{equation}
where $n=\pm 1,\pm 2,\ldots$. Solving Eq. (\ref{eq10b}) we get the following value for $\epsilon$
\begin{equation}
\epsilon=\frac{h_cR^2}{Dn^2}.
\label{eq11}
\end{equation}
Using the fact that $\epsilon=1/(\lambda-1)$ we get the following value for $\lambda$
\begin{equation}
\lambda=1+\frac{Dn^2}{h_cR^2}
\label{eq12}
\end{equation}
Equation (\ref{eq12}) is in agreement with Eq. (\ref{eq03b}) when $v_{EP}^2=h_c^2/\kappa^2$ and $k=n/R$. Interestingly, Eq. (\ref{eq12}) is valid only when conditions (i) and (ii) are fulfilled.\\
Once we know the value of $\epsilon$ we can substitute in $a^2=\epsilon v_{EP}^2/Dh_c$ in order to get $a=\epsilon$, substituting this value into Eq. (\ref{eq10}) we have the conditions that have to be satisfied in order to have unbroken-$\mathcal{PT}$ symmetry at the exceptional point which are $0<\epsilon<1$ and $0<\epsilon<2(1-\sqrt{1-\epsilon^2})$, which gives us the following solution
\begin{equation}
\frac{4}{5}<\epsilon<1
\label{eq13}
\end{equation}
Equation (\ref{eq13}) is the main result of this study which states that two phase transitions take place at the exceptional point and depends only on the radii of the rotating rings. \\
Substituting $a=\epsilon$ in Eq. (\ref{eq09}) we get $\chi_+^2=\epsilon^2-1$ and $\chi_-^2=-1$, therefore $f_1(z)=A_1\cos(\sqrt{1-\epsilon^2}z)+B_1\cos(z)$ which means we have to choose $A_1=0$ in order to fulfill the periodicity condition. Substituting $f_1$ into Eq. (\ref{eq05}) we obtain the following ordinary differential equation for $f_2$:
\begin{equation}
\frac{d^2f_2}{dz^2}+\epsilon\frac{df_2}{dz}+f_2=-\epsilon B_1\cos(z).
\label{eq12a}
\end{equation}
The general solution for Eq. (\ref{eq12a}) is given by
\begin{equation}
f_2(z)=A_2e^{-z/2\epsilon}\cos\left(\sqrt{1-(\epsilon/2)^2}z+\phi\right)-B_1\sin(z),
\label{eq12b}
\end{equation}
where $A_2$, $B_1$, $\phi$ and $n$ are constants to be determined by the initial conditions. \\
If we impose the following initial conditions over the temperature profiles in  the rings 
\begin{equation}
T_1(x,0)=T_2(x,0)=T_0+A\cos(x/R)
\label{eq14a}
\end{equation}
we need to choose $n=1$ and $B_1=A$ in order to get 
\begin{equation}
f_1(x)=A\cos(x/R)
\label{eq15}
\end{equation}
and 
\begin{equation}
f_2(x)=Ae^{-Dx/2R^3h_c}\sec(\phi)\cos\left(\alpha\frac{x}{R}+\phi\right)-A\sin(x/R)
\label{eq16}
\end{equation}
where $\alpha=\sqrt{1-(h_cR^2/2D)^2}$ and
\begin{equation}
\phi=\arctan\left[\cot(2\pi\alpha)-\csc(2\pi\alpha)e^{\pi D/h_cR^2}\right].
\label{eq17}
\end{equation}
The solution given in Eq. (\ref{eq15}) means that $\Delta T_1(x,0)=f_1(x)$, therefore the temperature distribution in the first ring will not change in position but will only decay on time, in contrast with the solution given in Eq. (\ref{eq16}) which is different from the initial condition, therefore the temperature profile will change in position and decay on time. In Fig. (\ref{fig2}) we show the Rabi oscillations as a function of position for different values of $\epsilon$. \\ 
\begin{figure}[htb]
  \begin{center}
    \begin{tabular}{cc}
      \resizebox{40mm}{!}{\includegraphics{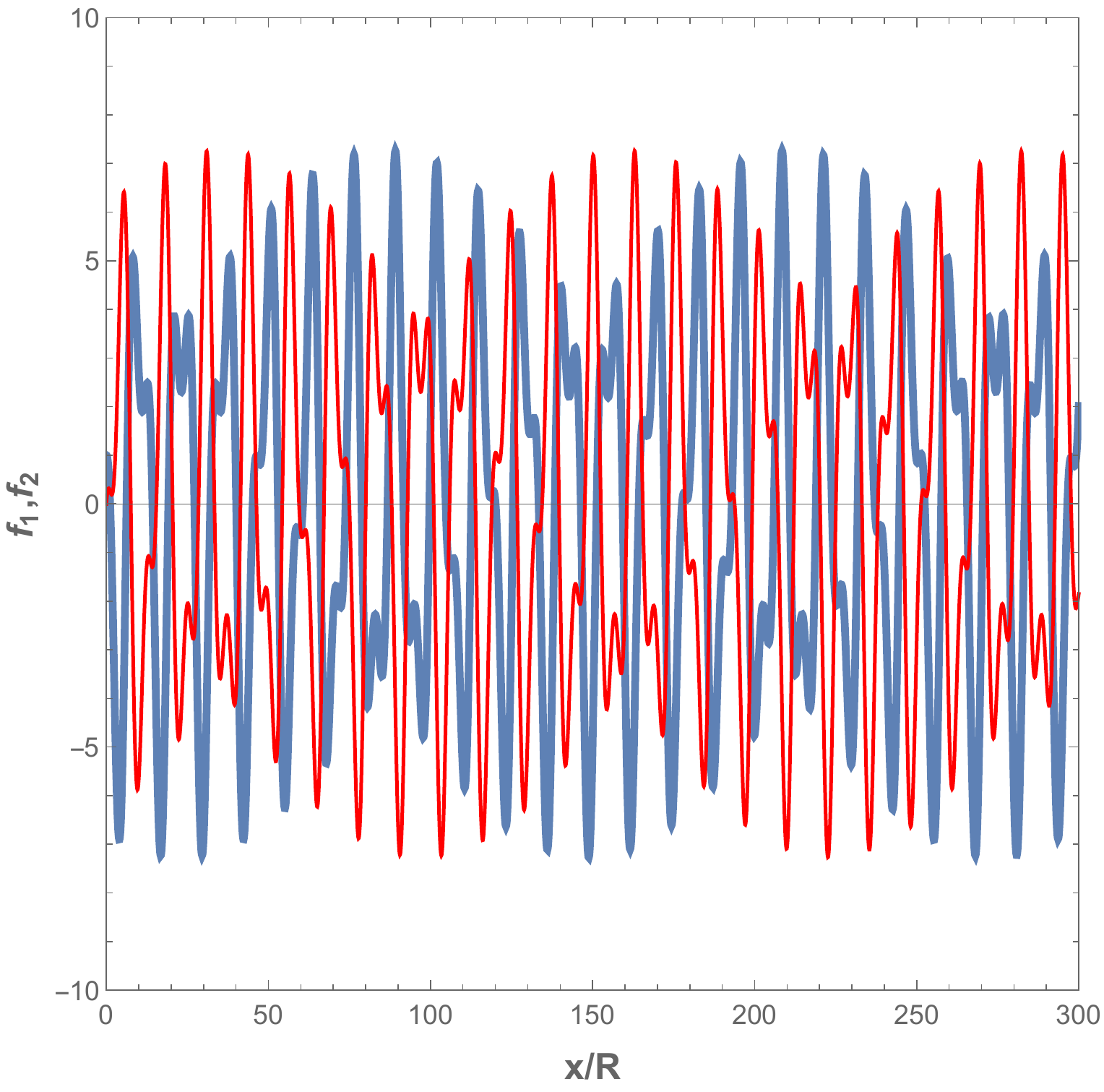}} &
      \resizebox{40mm}{!}{\includegraphics{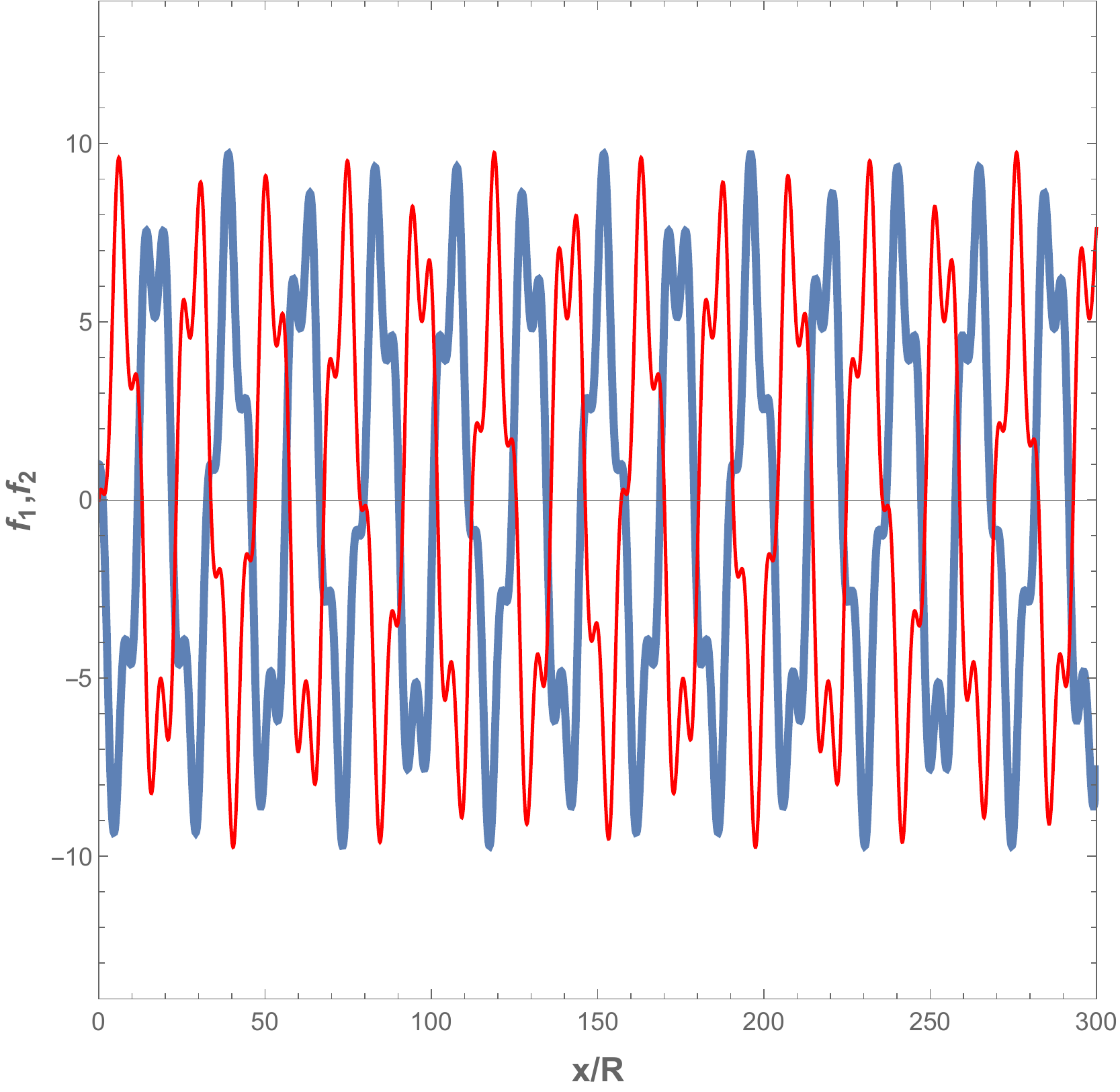}} \\
    \multicolumn{1}{c}{\mbox{\bf (a)}} &
\multicolumn{1}{c}{\mbox{\bf (b)}} \\
    \end{tabular}
    \caption{Numerical solution of the coupled equations given in Eq. (\ref{eq05}). The graph shows the Rabi oscillations for (a) $\epsilon=0.86$ and (b) $\epsilon=0.96$, respectively. The Rabi oscillations are harder to see near the upper end of the unbroken $\mathcal{PT}$ symmetry.}
\label{fig2}
  \end{center}
\end{figure}
If we impose the following new conditions over the temperature profiles in the rings
\begin{equation}
T_1(x,0)=T_0+A\cos(x/R) \quad \mbox{and} \quad T_2(x,0)=T_0+A\sin(x/R)
\label{eq18}
\end{equation}
we need to choose $n=1$, $B_1=-A$, $A_2=0$, which means that $\Delta T_i(x,0)=f_i(x)$, therefore both temperature distributions will remain invariant and will only decay on time.  \\
\begin{figure}[htb]
  \begin{center}
    \begin{tabular}{cc}
      \resizebox{40mm}{!}{\includegraphics{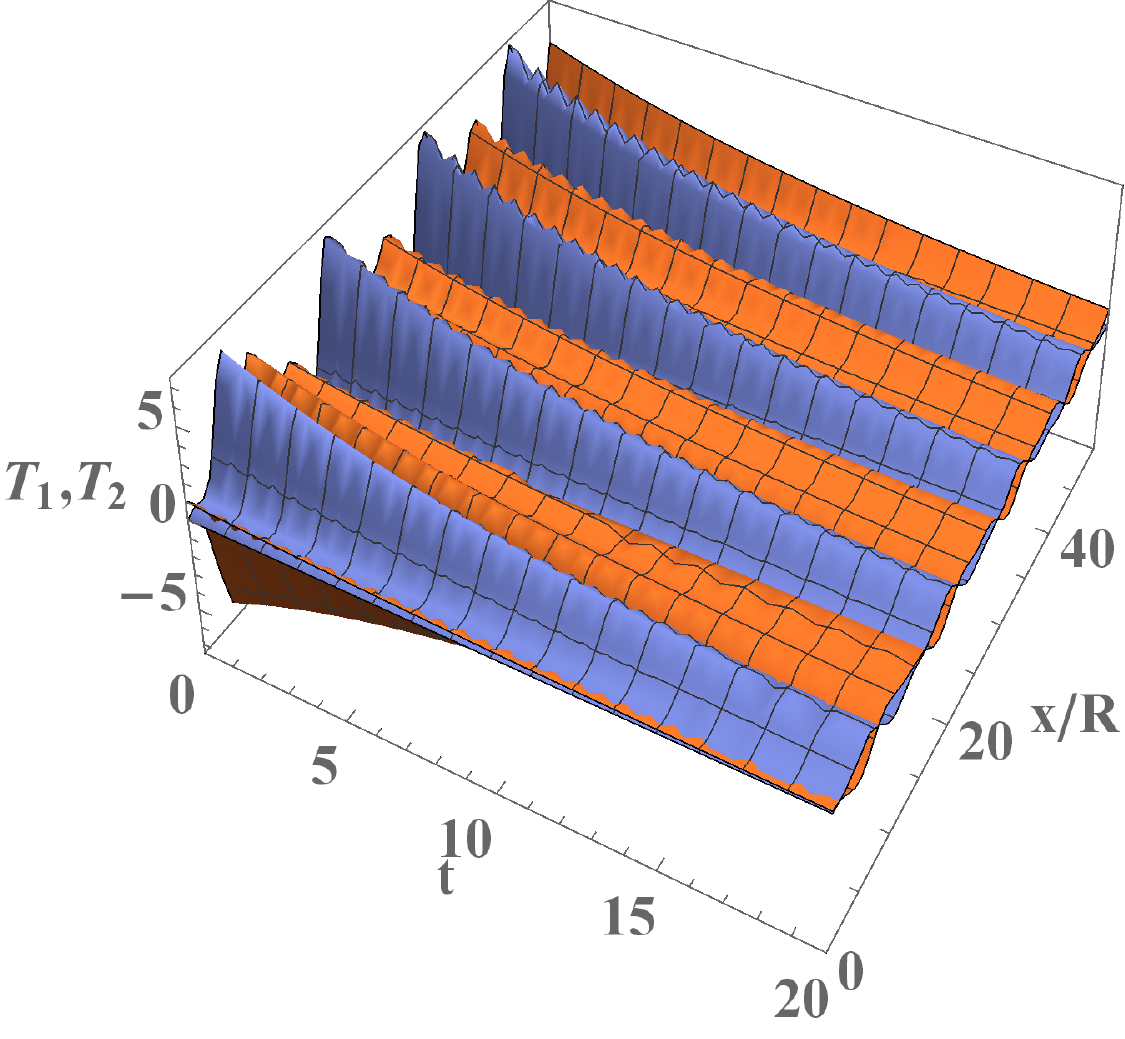}} &
      \resizebox{40mm}{!}{\includegraphics{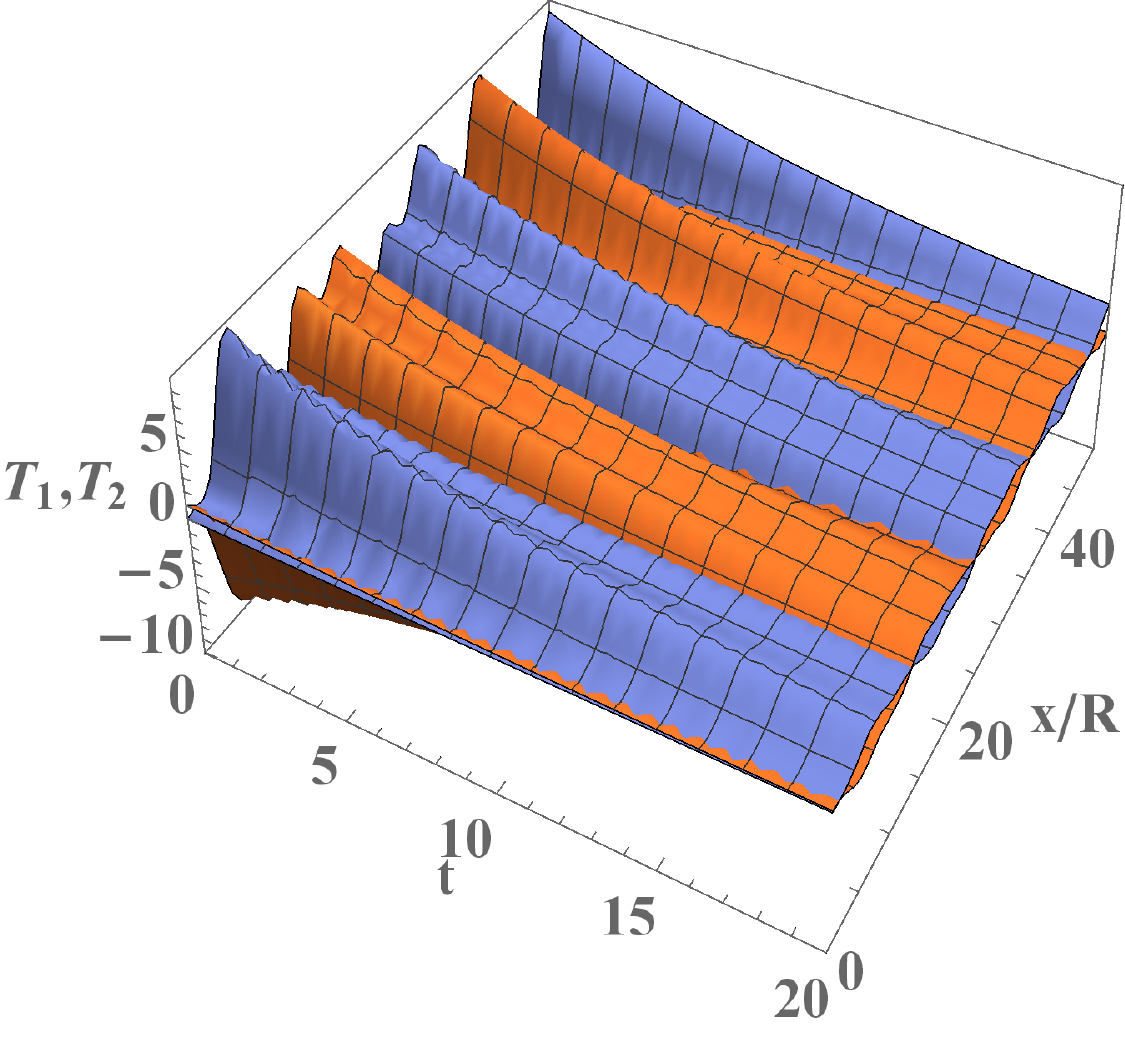}} \\
    \multicolumn{1}{c}{\mbox{\bf (a)}} &
\multicolumn{1}{c}{\mbox{\bf (b)}} \\
    \end{tabular}
    \caption{The graph shows the temperature profiles for both rings
    for the following values $D=100 mm^2/s$, $\rho=1000 Kg/m^3$, $c=1000 J/Kg ^{\circ}K$, $k_i=1 W/m ^{\circ}K$, $a=100 mm$, $b=5 mm$, $d=1 mm$ and for the radii: (a) $R=21 mm$ and $v_{EP}=4.2 mm/s$ and (b) $R=22 mm$ and $v_{EP}=4.4 mm/s$.}
\label{fig3}
  \end{center}
\end{figure}
Using the same experimental values given in Ref. \ref{Li}, i.e. $D=100 mm^2/s$, $\rho=1000 Kg/m^3$, $c=1000 J/Kg ^{\circ}K$, $k_i=1 W/m ^{\circ}K$, $a=100 mm$, $b=5 mm$ and $d=1 mm$, we find that Rabi oscillations take place for the fundamental wave if the inner ring radius is between $20 mm<R<22 mm$ and the rings are rotating with equal but opposite velocities given by $v_{EP}=h_cR$. In Fig. (\ref{fig3}) we show the temperature fields for the unbroken-$\mathcal{PT}$ regime where damped Rabi oscillations occur in which the maximum and minimum temperatures are 90$^{\circ}$ out of phase.  \\
Let us now consider the case when the rings are rotating with different velocities close to the exceptional point, specifically we will like to solve the following system
\begin{align}
\frac{\partial T_1}{\partial t}=D\frac{\partial^2 T_1}{\partial x^2}-\left(v_{EP}+\delta v\right)\frac{\partial T_1}{\partial x}+h_c(T_2-T_1) \nonumber \\
\frac{\partial T_2}{\partial t}=D\frac{\partial^2 T_2}{\partial x^2}+\left(v_{EP}-\delta v\right)\frac{\partial T_2}{\partial x}+h_c(T_1-T_2)
\label{eq19}
\end{align}
where $\delta v<<v_{EP}$. At first it seems that the system given in Eq. (\ref{eq19}) is not APT symmetric, however if we make the following transformation $\xi=x-\delta v t$ we obtain a system of equations identical to the one given in Eq. (\ref{eq03}) replacing $x\rightarrow\xi$ and $v\rightarrow v_{EP}$. The solution for Eq. (\ref{eq19}) is given by 
\begin{align}
\Delta T_1=e^{-\lambda h_c t}\cos((x-\delta v t)/R) \\ 
\Delta T_2=-e^{-\lambda h_c t}\sin((x-\delta v t)/R) 
\label{eq20}
\end{align}
which means that the temperature profiles are moving. This result shows that we can have a rest-to-motion temperature profile without having equal opposite rotating velocities. \\ 
In summary, we have predicted the existence of Rabi oscillations at the exceptional point in the diffusive system proposed by Li {\it et al.}. We showed that at the exceptional point the system exhibits two $\mathcal{PT}$ phase transitions which takes place at critical values for the radii of the rotating rings. Specifically, if the rings are rotating in opposite directions with equal tangential velocity given by $v_{EP}=h_c/|\kappa|$ and the ring radius lies between $\sqrt{4D/5h_c}<R<\sqrt{D/h_c}$ for the fundamental wave, i.e. $\kappa=\pm 1/R$, the temperature fields exhibit damped Rabi oscillations. We have shown also that it is not essential to have identical opposite rotating velocities for the rings in order to have a rest-to-motion temperature transition, we can do this also by increasing/decreasing the upper/lower ring velocity away from the exceptional point in order to obtain traveling wave solutions with positive/negative velocity.
Our work reveals the rich structure of exceptional points in anti-parity-time symmetric diffusive systems.\\ \\
I would like to acknowledge support by the program Cátedras Conacyt through project 1757 and from project A1-S-43579 of SEP-CONACYT Ciencia Básica and Laboratorio Nacional de Ciencia y Tecnología de Terahertz.

\end{document}